\documentclass[12pt]{article}
\begin{document}
\vspace*{-.6in}
\thispagestyle{empty}
\begin{flushright}
CALT-68-2234\\
hep-th/9908091
\end{flushright}
\baselineskip = 18pt

\vspace{.5in}
{\Large
\begin{center}
Remarks on Non-BPS D-Branes\footnote{Work
supported in part by the U.S. Dept. of Energy under Grant No.
DE-FG03-92-ER40701.}
\end{center}}

\begin{center}
John H. Schwarz\\
\emph{California Institute of Technology, Pasadena, CA  91125, USA}
\end{center}
\vspace{1in}

\begin{center}
\textbf{Abstract}
\end{center}
\begin{quotation}
\noindent
Following Sen's discovery of various stable non-BPS D-branes, K-theory has been
shown to be the appropriate mathematical framework for classifying conserved D-brane
charges. The classification accounts for known D-branes and predicts some new ones
including a D8-brane in type I superstring theory. After briefly reviewing these developments,
we discuss certain issues pertaining to the D8-brane, which is unstable.
\end{quotation}

\vfil
\centerline{\it Presented at Strings 99}

\newpage

\pagenumbering{arabic}

\section{Unstable Type II D-Branes}

During the past two years Ashoke Sen has pioneered the study of
non-BPS D-brane systems.  (For reviews see \cite{sen,lerda}.) In
particular, he has focused on systems of coincident D-branes and
anti-D-branes.  The basic idea is that whereas a system of
coincident D-branes (or anti-D-branes) would be a stable
supersymmetric (BPS) configuration, a system with both branes and
anti-branes is not.  Each separately preserves half of the
supersymmetries of the ambient background, but different halves
are preserved in each case, so that when both are present, there
is no unbroken supersymmetry.  One manifestation of this fact is
that the excitation spectrum of open strings connecting ${\rm
D}p$-branes to $\overline{{\rm D}p}$-branes has the reversed GSO
projection compared to ones connecting ${\rm D}p$-branes to ${\rm
D}p$-branes (or $\overline{{\rm D}p}$-branes to $\overline{{\rm
D}p}$-branes).  This results in tachyon fields on the world
volume, which signal an instability.  When the tachyon fields roll
to a minimum --- in a Higgs-like manner --- this represents
annihilation of branes and anti-branes.

To be specific, consider a system of $N$ ${\rm D}p$-branes and $N'$
$\overline{{\rm D}p}$-branes all of which are coincident $(p + 1)$-dimensional
hyperplanes embedded in ${\bf R}^{10}$.
The ground state of the ${\rm D}p$ - ${{\rm D}p}$ open strings
gives $U(N)$ gauge fields $A$ and the $\overline{{\rm D}p}$ - $\overline{{\rm D}p}$
open strings give
$U(N')$ gauge fields $A'$.  The ${\rm D}p$ - $\overline{{\rm D}p}$
open strings, on the other hand,
give a bifundamental $(N, N')$ tachyon $T$.  These fields can be written
together as a ``superconnection''
\begin{equation}
{\mathcal A} = \left(\begin{array}{cc}
A & T\\
\bar T & A'\end{array} \right).
\end{equation}
One issue is whether or not the branes and anti-branes can completely
annihilate.  The criterion, basically, is whether the total D-brane charge
(which is conserved) is zero or not.  Cancelling the ${\rm D}p$-brane charge requires
$N = N'$, of course, but that is not the whole story.  It is also necessary
that the gauge bundles $E$ and $E'$ (associated to branes and anti-branes)
should be topologically equivalent, $E \sim E'$.  Otherwise, there is some
lower-dimension D-brane charge, and such a D-brane would survive.  To
illustrate this consider the case of one ${\rm D}2$-brane and one
$\overline{{\rm D}2}$-brane, which are wrapped on a $T^2$, and coincident in the
other dimensions.  The Wess-Zumino term of the ${\rm D}2$-brane world-volume action
contains
\begin{equation}
\int (C e^F)_3 = \int_{R \times T^{2}} (C_3 + C_1\wedge F),
\end{equation}
where the $C$'s are $RR$ potentials.  From this formula we see
that the magnetic flux $\int_{T^{2}} F$ is a source of $C_1$,
which means that it carries ${\rm D}0$-brane charge.  Thus, for
example, if the ${\rm D}2$-brane has flux giving one unit of ${\rm
D}0$-brane charge and the $\overline{{\rm D}2}$-brane has no such
flux, then the annihilation leaves a ${\rm D}0$-brane
\begin{equation}
{\rm D}2 + \overline{{\rm D}2} \rightarrow {\rm D}0.
\end{equation}
The world-volume theory that describes a coincident ${\rm D}p + \overline{{\rm D}p}$ system
can be formulated in terms of the gauge fields and tachyons, where one imagines
that all other modes have been integrated out.  It is hard to make this
explicit in a controlled manner, since the tachyon mass is generally string
scale.  Thus the discussion that follows is necessarily somewhat qualitative
and heuristic.  It does have the advantage of being very physical and
intuitive, however.  Analyses with better mathematical control lead to the same
conclusions.  One approach is to use conformal field theory methods, as
described in Sen's lecture.  Another one is to use boundary-state techniques,
as described in Gaberdiel's lecture.  In any case, working with gauge fields
and tachyons, the world-volume theory has a tachyon potential $V(T)$, which
must be invariant under the $U(N) \times U(N')$ gauge symmetry.  Moreover, when
$N=N'$, Sen argues that it should have minima that correspond to pure vacuum.
The locus of minima, all of which are gauge equivalent, is given by $T = T_0\,
\mathcal{U}$, where $T_0$ is a fixed positive real number and $\mathcal{U}$ is
an arbitrary constant element of $U(N)$.  At the minimum, the tachyon
condensation energy should exactly cancel the energy of the D-branes
\begin{equation}
V(T_0\,  \mathcal{U}) + 2N T_{Dp} = 0.
\end{equation}
Here, $T_{{\rm D}p}$ is the tension of a single ${\rm D}p$-brane.  Thus when $E \sim E'$
and $T = T_0\, \mathcal{U}$, the ${\rm D}p + \overline{{\rm D}p}$ system is equivalent to
pure vacuum.  What happens to the $U(N)$ gauge groups is not completely
understood.

Let us now take $N = N'= 1$ and consider a kink configuration of the tachyon
field $T$.  $T$ is complex, so let us consider ${\rm Im} \, T = 0$ and ${\rm Re} \, T = T_0 \tanh
(x/a)$, where $x$ is one of the Cartesian coordinates on the branes.  This
describes a solitonic ${\rm D}(p - 1)$-brane of thickness $a$ concentrated in the
vicinity of $x = 0$.  (The precise functional form is not important.)  Since
the vacuum manifold $|T| = T_0$ is a circle, and $\pi_0 (S^1)$ is trivial, this
D-brane has a real tachyon in its world volume and is unstable.  This is just
as well, since the stable D-branes of type II theories are believed to be
known, and this one is not in the list.  In fact, such unstable D-branes can be
constructed for all ``wrong'' values of $p$ in type II theories.  Stable
D-branes exist for $p$ = even in the IIA theory and $p$ = odd in the IIB theory.
The unstable ones occur for the other values of $p$.  Sen has demonstrated that
these unstable D-branes are useful for analyzing certain issues.  My purpose in
describing them here is to set the stage for an analogous construction, which
will appear later.

\section{Non-BPS Type I D0-Branes}

Let me now review one of Sen's constructions of a non-BPS stable ${\rm D}0$-brane in
type I superstring theory.  The construction we will consider is in terms a
tachyon kink in a D-string anti-D-string configuration.  Recall that the type I
D-string is actually the Spin (32)/${\bf Z}_2$ heterotic string continued to strong
coupling.  The continuation is reliable, because the string is BPS.  A system
of $N$ coincident $D$ strings has world volume gauge group $O(N)$.  This can be
understood as the subgroup of $U(N)$ on a set of type IIB D-strings that survives
orientifold projection.  In particular, for a single D-string the group is
$O(1) = {\bf Z}_2$.  Even though there are no gauge fields in this case, the group
matters.  In particular, a D-string wrapped on a circular spatial dimension has
possible Wilson lines $W = \pm 1$.

The 32 left-moving fermion fields $\lambda^A$ on the D-string
world-sheet arise as zero modes of {\rm D}1 - {\rm D}9 open
strings.  When wrapped on a circular dimension, the Wilson line
encodes their periodicity
\begin{equation}
\lambda^A (x + 2\pi R) = W\lambda^A (x).
\end{equation}
Thus, for $W = 1$,  $\lambda^A$ has zero modes, which satisfy a Clifford algebra,
and D-string quantum states are gauge group spinors (with $2^{15}$ components).

Now consider a ${\rm D}1 + \overline{{\rm D}1}$ pair wrapped on a circle.  If one string
has $W = 1$ and the other one has $W = - 1$, then the overall two-particle
state is a gauge group spinor.  Since the gauge group is not broken, this
implies that complete annihilation is not possible.  The tachyonic ground state of the
open string connecting the D-string and the anti-D-string is real in this case.
For the case of opposite Wilson lines that we are considering, the tachyon field is
antiperiodic.  Thus it has the Fourier series expansion
\begin{equation}
T = \sum_n T_{n + 1/2} (t) \exp \left[i \left(\frac{n +
1/2}{R}\right) x \right].
\end{equation}
The mass of $T_{n + 1/2}$, considered as a particle in $9d$, is
\begin{equation}
M_{n + 1/2}^2 = (n + 1/2)^2/R^2 - 1/2.
\end{equation}
The $-1/2$ term is the tachyonic mass-squared value (in string
units) in 10d, as usual for an RNS string.  From this formula we
see that for $R <1/\sqrt{2}$, there is no tachyonic instability
and the wrapped ${\rm D}1 + \overline{{\rm D}1}$ pair does not
annihilate.  For $R > 1/\sqrt{2}$, on the other hand, $T_{\pm
1/2}$ (and possibly other modes) are tachyonic.  This means that
the strings can annihilate. What results is a stable non-BPS ${\rm
D}0$-brane, which is a gauge group spinor.  It carries a conserved
${\bf Z}_2$ charge.  In this case, the ${\bf Z}_2$ corresponds to
the two conjugacy classes of Spin (32)/${\bf Z}_2$.

At $R = R_c = 1/\sqrt{2}$ and small string coupling constant $g$
\begin{equation}
M_{{\rm D}0} \sim 2 \cdot 2\pi R_c \cdot T_{{\rm D}1} = \sqrt{2}/g.
\end{equation}
Sen has argued that this is the leading small $g$ value of the
type 1 ${\rm D}0$-brane mass for all $R$, though there are
higher-order corrections.  It has the usual $1/g$ factor that is
characteristic of D-branes.  Curiously, its mass differs from that
of the type IIA ${\rm D}0$-brane by a factor of $\sqrt{2}$ (in
leading order).  In the S-dual heterotic theory the lightest gauge
group spinor occurs at the first excited level in the perturbative
spectrum.  Presumably, the non-BPS ${\rm D}0$-brane of type I is
this state continued to strong coupling.

\section{K-Theory Classification of D-Branes}

Recall that a ${\rm D}p + \overline{{\rm D}p}$ system is
characterized by a pair of vector bundles $(E, E')$ and a tachyon
$T$, which is a section of $E^{\star} \otimes E'$. Complete
annihilation should be possible if and only if $E \sim E'$.  This
requires $N = N'$, in particular.  Following an earlier suggestion
by Moore and Minasian \cite{moore}, Witten has argued that
equivalence classes of pairs $(E, E')$ that can be related by
brane annihilation and creation correspond to K-theory classes
\cite{witten}.  So these are the mathematical objects that
correspond to conserved D-brane charges.

For example, D-brane charges of the type IIB theory on ${\bf R}^{10}$ are given by
\begin{equation}
\tilde{K} (S^{9-p}) = \left\{\begin{array}{ll}
{\bf Z} & p = {\rm odd}\\
0 & p = {\rm even}\end{array}\right. .
\end{equation}
This accounts for the RR charges of all stable type IIB D-branes.  Note that
the unstable D-branes (for $p$ = even) carry no conserved charges and do not
show up in this classification.

In the case of type I theory, $E$ is an $O(N + 32)$ bundle and $E'$ is an
$O(N)$ bundle, so that the total RR 9-brane charge is 32.  The relevant
K-theory groups for ${\bf R}^{10}$ in this case are denoted
$\widetilde{KO}(S^{9-p})$, as explained by Witten.

\noindent The results are as follows:

\begin{itemize}
\item $\widetilde{KO} (S^{9-p}) = {\bf Z}$ for $p = 1, 5, 9$

\noindent these classify the charges for the three kinds of BPS Dp-branes of
type I.

\item $\widetilde{KO} (S^{9-p}) = {\bf Z}_2$ for $p = - 1, 0, 7, 8$

\noindent $p =  -1$ corresponds to the type I D-instanton, and $p = 0$
corresponds to the non-BPS ${\rm D}0$-brane, which we have discussed.  The cases $p =
7,8$ are additional non-BPS D-branes proposed by Witten.

\item $\widetilde{KO} (S^{9-p}) = 0$ for $p = 2,3,4,6$

\noindent there are no conserved D-brane charges in these cases.
\end{itemize}

\section{Issues Concerning the Type I D8-brane}

The K theory classification of type I D-branes, which we have just summarized,
suggests two new D-branes not discussed previously: ${\rm D}7$ and ${\rm D}8$, each of
which is supposed to carry a conserved ${\bf Z}_2$ charge.

As noted in the final paragraph of Ref.~\cite{frau}, there is a
tachyon in the spectrum of {\rm D}7 - {\rm D}9 and {\rm D}8 - {\rm
D}9 open strings. This means that the world volume of a ${\rm
D}7$-brane or ${\rm D}8$-brane contains 32 tachyon fields.
Therefore neither of these D-branes is stable.  This raises the
question of what happens to the conserved ${\bf Z}_2$ charge when
they dissolve into the background ${\rm D}9$-branes.  The comments
that follow arose out of discussions with Oren Bergman and Ashoke
Sen, as well as correspondence with Edward Witten. I will only
discuss the ${\rm D}8$-brane, though the ${\rm D}7$-brane story is
likely to be similar.

Witten has argued in support of the ${\rm D}8$-brane as follows:
The type I D-instanton implies that there are two different
``vacua'', distinguished by the sign of the instanton amplitudes.
This is a $Z_2$ analog of the $\theta$ angle in QCD.  One should
expect that there is a domain wall separating the two vacua and
this should be the ${\rm D}8$-brane.  The sign change of instanton
amplitudes would mean that the D-instanton is the EM dual of the
${\rm D}8$-brane. Investigations that support this picture were
carried out by Gukov~\cite{gukov}.

The K-theory analysis incorporates Bott periodicity.  This
suggests that the type I ${\rm D}8$-brane should have features in
common with the type I {\rm D}0-brane, discussed in Section 2. Of
course, Bott periodicity should be taken {\it cum grano salis},
since the total spacetime dimension is ten.  The construction of
the type I ${\rm D}0$-brane that we described involved wrapping
D-strings on a circle, which was a convenient regulator.  However,
one might argue that a localized ${\rm D}8$-brane should not exist
on a circle (in the direction normal to the brane), since this
would require identifying the two distinct vacua.  Therefore we
will analyze the situation in uncompactified ${\bf R}^{10}$.

The ${\rm D}0$-brane could have been presented without involving
compactification. In any case, by considering the $R \rightarrow
\infty$ limit of the construction in Section 2, we see that the
${\rm D}0$-brane can be described as a tachyonic kink in a system
consisting of an infinite straight D-string and a coincident
anti-D-string.  The kink configuration would be exactly the same
as we described for type II theories in Section 1.  However,
unlike the type II examples, the tachyon field is real in this
case, and the potential $V(T)$ is an even function because of the
${\bf Z}_2$ gauge symmetries.  The kink configuration describing
the ${\rm D}0$-brane is topologically stable in this case because
the vacuum manifold is $S^0$ ($T=\pm T_0$) and $\pi_0 (S^0) = {\bf
Z}_2$.

Let us now try to construct the ${\rm D}8$-brane out of ${\rm
D}9$-branes in an analogous manner.  One essential difference is
that the total ${\rm D}9$-brane charge must be 32.  Therefore the
simplest analog to consider is 33 ${\rm D}9$-branes and one
$\overline{{\rm D}9}$-brane filling the entire ${\bf R}^{10}$
spacetime.  In this case the open strings connecting the
$\overline{{\rm D}9}$ to the ${\rm D}9$-branes give 33 real
tachyon fields $\overrightarrow{T}$ in the fundamental
representation of $SO(33)$. (It doesn't matter whether one uses
$O(N)$ or $SO(N)$ in the present setting.)  The potential
$V(\overrightarrow{T})$ must have $SO(33)$ symmetry and therefore
the vacuum manifold should be given by $|\overrightarrow{T}| =
T_0$, which describes an $S^{32}$. This manifold is connected, so
there is no topologically stable kink. This is the same situation
we encountered for the unstable type II D-branes in Section 1.  In
this case there are 32 directions of instability, so one expects
to find 32 tachyon fields in the ${\rm D}8$-brane world volume.
This agrees with the conclusion of Ref.~\cite{frau}, which
identified them with modes of {\rm D}8 - {\rm D}9 open strings.

So what are we to make of all this?  I think it is clear that the
${\rm D}8$-brane is unstable, at least unless something further is
done to stabilize it.  Still, it may be interesting to consider
setting up a ${\rm D}8$-brane configuration and exploring what
that implies.  I won't present the details of the reasoning here,
but it appears that the vacua on the two sides of the ${\rm
D}8$-brane are distinguished by the chirality of gauge group
spinors.  Of course, once the ${\rm D}8$-brane decays, eventually
leaving a uniform type I vacuum, only one chirality will remain.
This may sound paradoxical, but it is possible because the gauge
group is broken inside the ${\rm D}8$-brane.

In conclusion, K-theory classifies D-brane charges.  However, high
dimension non-BPS D-branes are sometimes destabilized by tachyonic
modes of open strings connecting them to background spacetime
filling D-branes.

\section*{Acknowledgments}

I am grateful to Oren Bergman, Ashoke Sen, and Edward Witten for helpful
discussions and suggestions.


\newpage

\begin{thebibliography}{9}

\bibitem{sen} A. Sen, ``Non-BPS States and Branes in String Theory,'' hep-th/9904207.

\bibitem{lerda} A.Lerda and R. Russo, ``Stable Non-BPS States in String Theory:
a Pedagogical Review,'' hep-th/9905006.

\bibitem{moore} G. Moore and R. Minasian, ``K-Theory and Ramond-Ramond Charges,''
{\bf JHEP} 9711:002, 1997, hep-th/9710230.

\bibitem{witten} E. Witten, ``D-Branes and K-Theory'', hep-th/9810188.

\bibitem{frau} M. Frau, L. Gallot, A. Lerda, and P. Strigazzi,  ``Stable Non-BPS D-Branes
in Type I theory,'' hep-th/9903123.

\bibitem{gukov} S. Gukov, ``K-Theory, Reality, and Orientifolds,'' hep-th/9901042.

\end{thebibliography}
\end{document}